\documentclass[epj]{webofc}
\usepackage[varg]{txfonts}   
\usepackage[usenames]{color}

\newcommand{\bp}{{\bf p}}

\newcommand{\bX}{{\bf X}}

\newcommand{\cA}{{\mathcal A}}
\newcommand{\cB}{{\mathcal B}}
\newcommand{\cC}{{\mathcal C}}

\wocname{Mathematical Modeling and Computational Physics}
\woctitle{Mathematical Modeling and Computational Physics 2015}

\usepackage[english]{babel}
\begin{document}
\title{The stumbling block of the Gibbs entropy: the reality of the negative absolute temperatures}
%
%

\author{Drago\c s-Victor Anghel\inst{1}\fnsep\thanks{\email{dragos@theory.nipne.ro}}
}

\institute{Institutul National de Fizica si Inginerie Nucleara--``Horia Hulubei''}

\abstract{%
  %
  The second Tisza-Callen postulate of equilibrium thermodynamics states that for any system exists a function of the system's extensive parameters, called \textit{entropy}, defined for all equilibrium states and having the property that the values assumed by the extensive parameters in the absence of a constraint are those that maximize the entropy over the manifold of constrained equilibrium states.
  By analyzing the evolution of systems of positive and negative absolute temperatures, we show that this postulate is satisfied by the Boltzmann formula for the entropy and is violated by the Gibbs formula. Therefore the Gibbs formula is not a generally valid expression for the entropy.
  %
  
  Viceversa, if we assume, by \textit{reductio ad absurdum}, that for some thermodynamic systems the equilibrium state is determined by the Gibbs' prescription and not by Boltzmann's, this implies that such systems have macroscopic fluctuations and therefore do not reach thermodynamic equilibrium.
}
\maketitle
\section{Introduction} \label{intro}


Recently a heated debate sparkled about the correct expression for the entropy of a system (see Refs. \cite{Nat.Phys.10.67.2014.Dunkel,PhysRevE.90.062116.2014.Hilbert,PhysRevE.91.052147.2015.Campisi,AmJPhys.83.163.2015.Frenkel,PhysRevE.92.020103.2015.Swendsen} and citations therein). From a variety of proposals, two expressions stand out: the Gibbs entropy,
\begin{subequations} \label{def_SBG}
\begin{equation}
  S_G = k_B \ln \Omega(E,\bX) , \label{def_SG}
\end{equation}
and the Boltzmann entropy,
\begin{equation}
  S_B = k_B \ln \big[ \omega(E,\bX) \epsilon \big] . \label{def_SB}
\end{equation}
\end{subequations}
In Eqs. (\ref{def_SBG}) $E$ is the energy of the system, $\bX \equiv (X_1, \ldots, X_n)$ is the collection of extensive external parameters (other than energy) that specify the state of the system, and $\epsilon$ is an arbitrary, small parameter with dimensions of energy. $\Omega$ represents the number of states of the system with energy less or equal to $E$ (at fixed $\bX$), whereas $\omega$ is the density of states (DOS), so that $\Omega \equiv \int_0^E \omega(E')\,dE'$. For a quantum system, if $H(\xi; \bX)$ is the Hamiltonian and $\xi$ are the microscopic degrees of freedom, then
\begin{equation}
  \Omega(E; \bX) \equiv {\rm Tr} \Theta[E-H(\xi; \bX)] \quad {\rm and} \quad 
  \omega(E; \bX) \equiv {\rm Tr} \delta[E-H(\xi; \bX)] = \frac{\partial \Omega(E, \bX)}{\partial E} . \label{def_omegas}
\end{equation}
In general, for thermodynamic systems with unbounded energy and monotonic DOS, the thermodynamic predictions of the two expressions (\ref{def_SBG}) coincide. Disagreements appear for mesoscopic systems and in systems with non-monotonic DOS. In the latter case the definition (\ref{def_SB}) may lead to negative temperatures whereas the temperatures derived from Eq. (\ref{def_SG}) are always positive.

\section{Thermodynamics} \label{sec_thermodynamics}


We have first to clarify the thermodynamic premises.
The state of a system is defined by the small set of parameters $(E,\bX)$. The external parameters $\bX$ may be directly measured, but the internal energy $E$ is defined by the work done in adiabatic processes (see for example \cite{Callen:book, Lungu:book, JMathChem.28.313.2000.Pogliani}).

The basic ingredient in any thermodynamic considerations is the existence of \textit{equilibrium states}. We assume that at constant external parameters and fixed internal constrains each system is either in equilibrium or evolves irreversibly towards an equilibrium state.
The equilibrium or the evolution towards equilibrium may be observed only in macroscopic systems by measurements that (supposedly) do not influence the set of parameters $(E,\bX)$.
In a mesoscopic system (a system in which the finite size effects are non-negligible) the fluctuations are observable or comparable to the measured quantities. \textit{In this sense} the equilibrium cannot be attained or the act of observing the equilibrium might perturb the state of the system.
Therefore we understand the equilibrium (in macroscopic and mesoscopic systems) as \textit{the state which is attained after a long interval of time ($t\to\infty$) when all the external parameters and internal constraints are fixed}. With this definition, any system (macroscopic or mesoscopic) is in equilibrium or tends to equilibrium when the external conditions are fixed.

We shall say that two or more systems are in \textit{thermal contact} if they can exchange energy without changing the parameters $\bX$. If the net (average) energy exchange between two systems in thermal contact is zero when the parameters $\bX$ are fixed, we say that the two systems are in \textit{thermal equilibrium}. (We do not discuss here the important issue of how the heat exchange may be observed, especially in a mesoscopic system.) Obviously, two identical copies of a system are in thermal equilibrium with one another (in the absence of any external forces that should break the symmetry between them).

The relation of thermal equilibrium (which we shall denote by ``$\sim$'') is \textit{reflexive} ($\cA \sim \cA$) and \textit{symmetric} ($\cA \sim \cB \ \Rightarrow \ \cB \sim \cA$).
If ``$\sim$'' is also \textit{transitive} (i.e. $\cA \sim \cB$ and $\cB\sim\cC$ imply $\cA \sim \cC$), then the thermal equilibrium is an \textit{equivalence relation}.

If we assume that \textit{the thermal equilibrium is an equivalence relation}, then the equivalence classes of a system form a set of disjoint \textit{isothermal} sets in the $(n+1)$-dimensional [$(n+1)$D] space of parameters $(E,\bX)$. If these sets are $n$D hyper-surfaces, then we identify each isothermal surface by a number $\theta$. If we can define an order for these numbers, then we call $\theta$ the \textit{empirical temperature}.

To choose the order of $\theta$, let's assume that for each $\bX_0$ fixed, the line defined by the points $(E,\bX_0)$, parallel to the $E$ axis, intersects each isothermal hyper-surface in only one point. Then we may (typically) order $\theta$ on intervals in increasing order of $E$ ($\theta$ does not have to be positive). Since the isothermal hyper-surfaces are disjoint, the order of $\theta$'s is independent of $\bX_0$. Furthermore, if $\theta(E,\bX_0)$ is a bijective function of $E$, then one may invert it to $E(\theta,\bX_0)$ and define the state of the system by $(\theta,\bX_0)$ instead of $(E,\bX_0)$.

Let us now discuss the existence of the \textit{empirical entropy}. For this we assume that one can perform quasistatic, reversible, adiabatic processes on systems. Two states $\cA$ and $\cB$ that can be connected by such a process are denoted by $\cA \bowtie \cB$. Then, obviously, $\cA \bowtie \cA$ and, since the processes are reversible, $\cA \bowtie \cB$ implies $\cB \bowtie \cA$. Moreover, one can combine two reversible adiabatic processes $\cA \bowtie \cB$ and $\cB \bowtie \cC$ to obtain $\cA \bowtie \cC$. This implies that ``$\bowtie$'' is an equivalence relation on the set of equilibrium states of the system and defines equivalence classes. The equivalence classes are called \textit{isentropic}.

We employ the \textit{Carath\'eodory formulation of the second principle of thermodynamics} \cite{Lungu:book, JMathChem.28.313.2000.Pogliani}, i.e. ``in the neighborhood of any equilibrium state of a system (of any number of thermodynamic coordinates), there exist states that are inaccessible by reversible adiabatic processes.'' If the isentropic sets are disjoint $n$D hyper-surfaces--like the isotherms--and the lines defined by the points $(E,\bX_0)$ (where $\bX_0$ is fixed) intersect them in only one point, we can define the \textit{empirical entropy} $\sigma(E,\bX_0)$ as a monotonically increasing function of $E$, for each $\bX_0$. The entropy is a state function. Any entropy function should be a bijective transformation of $\sigma(E,\bX)$.

Effectively, the entropy and the temperature may be constructed using the definition of heat and its property of being a holomorphic Pfaff form:
\begin{equation}
  \delta Q/\theta = \delta \sigma , \label{def_heat}
\end{equation}
where $\delta Q$ is the heat exchange and $\delta \sigma$ is the corresponding variation of the entropy. If the heat transfer occurs at constant $\bX$, then $\delta Q = \delta E$ and we obtain from (\ref{def_heat}) the standard relation
\begin{equation}
  \partial \sigma / \partial E \equiv 1/\theta . \label{def_temp}
\end{equation}
Choosing the value $\theta_0 = \theta(E,\bX)$ for one isothermal hyper-surface $S_{\theta_0}$ and the value of $\sigma_0 = \sigma(E,\bX)$ for one isentropic hyper-surface $S_{\sigma_0}$, one can construct the function $\sigma(E,\bX)$ on $S_{\theta_0}$ by integrating over $Q/\theta_0$ along any path on $S_{\theta_0}$. Once we know $\sigma(E,\bX)$ on $S_{\theta_0}$, we can extend it in the whole space by using the equation for the isentropic hyper-surfaces. Furthermore, having $\sigma(E,\bX)$ in the whole space, one can construct $\theta(E,\bX)$ by using Eq. (\ref{def_temp}). In all this we assume that $\theta$ does not take the value zero anywhere and the isothermal and isentropic hyper-surfaces are smooth.

General values for $\theta$ can be obtained by defining a thermometer which may be used to probe the temperature in any system by using the transitivity property of the temperature. The typical absolute temperature scale is the Kelvin scale. Nevertheless, as we shall see further, this scale is not sufficient to define the temperature in any physical system. For some systems we need a scale which contains also negative temperatures.

\subsection{Tisza-Callen postulates} \label{subsec_TC_postulates}

We saw very briefly how one can obtain some of the basic properties of thermodynamic systems, including the existence of temperature and entropy, starting from very general assumptions and without making reference to the principles of thermodynamics. Without going further into details we conclude the section by presenting the axiomatic foundation of thermodynamics of Tisza and Callen (see for example \cite{Callen:book, JNonNewtonianFluidMech.96.5.2001.Jongschaap, PhysRevE.92.020103.2015.Swendsen,Lungu:book}), which is based on the following four postulates (and includes the assumptions made above).

{\it Postulate 1 (existence of equilibrium states):
	Any isolated system has equilibrium states that are characterized uniquely by a small number of extensive variables $(E,\bX)$.}

{\it Postulate 2 (existence of entropy):
	There exists a function (called the entropy $S$) of the extensive parameters, defined for all equilibrium states and having the following property. The values assumed by the extensive parameters in the absence of a constraint are those that maximize the entropy over the manifold of constrained equilibrium states.}
We use here the notation $S$ for the entropy, to distinguish it from the empirical entropy $\sigma$.
This postulate is an expression of the \textit{second law} of thermodynamics. 

{\it Postulate 3 (additivity and differentiability of $S$):
	The entropy of a composite system is additive over the constituent subsystems (whence the entropy of each constituent system is a homogeneous first-order function of the extensive parameters). The entropy is continuous and differentiable. 
	}
This postulate applies only to systems in which the interaction between particles belonging to different sub-systems are negligible. Moreover, in the original formulation of Callen \cite{Callen:book} it is stated that $S$ increases monotonically with $E$, but this is unnecessary, as explained in \cite{PhysRevE.92.020103.2015.Swendsen}.

This postulate implies the existence of the temperature $T$, which is $T \equiv (\partial S/\partial E)^{-1}$. Using this definition and the maximization of the entropy of a composite system at equilibrium, we obtain that
\begin{equation}
  \frac{1}{T} = \frac{\partial S_1}{\partial E_1} = \frac{\partial S_2}{\partial E_2} = \ldots , \label{transitivity}
\end{equation}
hence the transitivity of the thermal equilibrium and the \textit{zeroth law} of thermodynamics \cite{PhysRevE.92.020103.2015.Swendsen}.

From this postulate we obtain also the \textit{first law} of thermodynamics, namely
\begin{equation}
  dE = T\,dS + \sum_{i=1}^n p_i \,d X_i \equiv \delta Q + \delta L \label{Lex1}
\end{equation}
where $\delta L$ is the work and $\bp$ is the collection of intensive variables conjugated to the variables $\bX$ \cite{PhysRevE.92.020103.2015.Swendsen}.

{\it Postulate 4:
	The entropy of any system vanishes in the state for which $T \equiv (\partial S/\partial E)^{-1} = 0$.
	}
This postulate expresses the \textit{third law} of thermodynamics, but shall ignore it in the following.

Now we can see if different definitions of the entropy comply with the axiomatic formulation of thermodynamics and compare the conclusions of Refs. \cite{Nat.Phys.10.67.2014.Dunkel,PhysRevE.90.062116.2014.Hilbert,PhysRevE.91.052147.2015.Campisi,AmJPhys.83.163.2015.Frenkel,PhysRevE.92.020103.2015.Swendsen}.

\section{The Gibbs entropy} \label{sec_G}

In searching for a microscopic expression for the entropy of a system which would be generally valid, some authors \cite{Nat.Phys.10.67.2014.Dunkel, PhysRevE.90.062116.2014.Hilbert, PhysRevE.91.052147.2015.Campisi} strongly support the Gibbs entropy (\ref{def_SG}).
Let's see if this satisfies the basic thermodynamic requirements outlined in Section~\ref{sec_thermodynamics} (except the \textit{postulate 4}).

Clearly, the existence of an equilibrium state is not contradicted by this definition and therefore \textit{postulate 1} is satisfied. Moreover, if two or more systems are in contact and some constraints are removed (like the removal of a wall or allowing heat exchange between different subparts of a system), the number of states $\Omega$ can only increase, since the states accessible before removing the constraints are still accessible after the removal (see for example Eqs. 48 and 49 of Ref. \cite{PhysRevE.90.062116.2014.Hilbert}). Apparently this implies that the \textit{postulate 2} is satisfied, since the entropy always increases after removal of some constraints. We shall come back to this point later and show that this is not the case.

The \textit{postulate 3} is satisfied only in the thermodynamic limit, since by putting in contact two systems one obtains a total system of entropy which is always bigger than the sum of the entropies of the isolated systems. \textit{Assuming} that the difference is negligible for thermodynamic systems, we can say that the \textit{postulate 3} is also satisfied in such cases \cite{PhysRevE.90.062116.2014.Hilbert}. 

Now let's see if the \textit{postulate 2} is indeed satisfied and $S_G$ also describes the equilibrium state of a composite system. For this we calculate the Gibbs temperature, defined by Eq. (\ref{transitivity}),
\begin{equation}
  T_G \equiv \frac{\Omega(E,\bX)}{k_B \omega(E,\bX)} . \label{def_TG}
\end{equation}
We take two relevant examples: the ideal gas and a system of independent spins in uniform magnetic field.

For an \textit{ideal gas} of $N$ particles in a volume $V$, the total number of states is $\Omega_{\rm id}(E, V, N) = C_{\rm id}(N) V^N E^{3N/2}$, where $C_{\rm id}(N)$ is a constant that depends only on $N$. Therefore the relation between the temperature and the energy of the system is $T_{G {\rm id}} = [2/(3 k_B)] E/N$, which takes values between $(2/3) \epsilon_{\rm min}/k_B$ (when $E\to E_{\rm min}$ and $\epsilon_{\rm min} \equiv E_{\rm min}/N$) and $\infty$ (when $E \to \infty$); $E_{\rm min}$ is the minimum energy of the system and in general is taken to be equal to zero.

The energy of a \textit{system of $N_0$ spins} ($N_0\gg 1$) in uniform magnetic field $B$ is
\begin{equation}
  E = - B\mu \sum_{i=1}^{N_0} s_i , \label{en_spins}
\end{equation}
where $\mu$ is the magnetic moment, $s_i = \pm 1/2$, is the spin orientation, and we assume for convenience that $B > 0$. The minimum energy is $E_0 = - B\mu N_0/2$, which is reached when all the spins are pointing upwards, and the maximum energy is $E_1 = B\mu N_0/2$ which is reached when all the spins are pointing downwards.
If we denote by $N$ the number of ``spin flips'' (which is the number of spins oriented downwards), then the energy of the system relative to $E_0$ is $E = B\mu N$ and the DOS is $\omega_s(N) \equiv \omega_s(E = B\mu N) = N_0!/[N! (N_0-N)!]/(B\mu)$ (see \cite{PhysRevE.91.052147.2015.Campisi} for details).
Obviously, $\omega_s(N)$ reaches its maximum when $N = N_0/2$ (when $N_0$ is even) or $N = (N_0-1)/2, (N_0+1)/2$ (when $N_0$ is odd). Since $N_0 \gg 1$, we shall say the maximum number of microconfigurations is $\omega_{s\, {\rm max}} = N_0!/[(N_0/2)!]^2/(B\mu)$. The total number of states is $\Omega_{s} (E) = B\mu \sum_{N=0}^{E/(B\mu)} \omega_s(N)$, and the Gibbs temperature $T_{G s} = \Omega_{s} (E)/[k_B \omega_s(E)]$ takes values between $B\mu/k_B$ (when $E \to 0$) and $(B\mu/k_B) \Omega_s(B\mu N_0)$ (when $E \to E_{\rm max} = B\mu N_0$).

Now we can see why $T_G$ is not a proper temperature and therefore $S_G$ is not an appropriate definition for the entropy. It is well known that if $E > B\mu N_0/2$, the spin system has a population inversion and cannot be in thermal equilibrium with an ideal gas. 
Yet, since $T_{G {\rm id}}$ takes values between $(2/3) \epsilon_{\rm min}/k_B$ and $\infty$ and if $2\epsilon_{\rm min}/3 < B\mu \Omega_s (B\mu N_0)$, then we can find the energies $E_{\rm id}$ and $E_s\ (> B\mu N_0/2)$, such that $T_{G {\rm id}}(E_{\rm id}) = T_{G s}(E_s)$.
Since the Gibbs temperatures are equal, the Gibbs entropy of the total system is maximized for these choices of energies, which correspond to a nonequilibrium state, so \textit{the Gibbs entropy is unphysical} (see also the next section).

\section{The Boltzmann entropy} \label{sec_B}

The Boltzmann entropy (\ref{def_SB}) is eventually the most used in statistics (see for example \cite{AmJPhys.83.163.2015.Frenkel,PhysRevE.92.020103.2015.Swendsen} related to the recent debate), although is very much criticized by the authors who consider the Gibbs entropy (\ref{def_SG}) as the only viable choice \cite{Nat.Phys.10.67.2014.Dunkel, PhysRevE.90.062116.2014.Hilbert, PhysRevE.91.052147.2015.Campisi}. The interpretation of $S_B$ is different than that of $S_G$ and it has been thoroughly discussed in Refs. \cite{AmJPhys.83.163.2015.Frenkel,PhysRevE.92.020103.2015.Swendsen}. 
Let's see if it satisfies the postulates.

It is easy to see that Eq. (\ref{def_SB}) is in accord with \textit{postulate 1}. 
Further, the evolution towards equilibrium is an evolution towards the maximum DOS. If we put two systems $\cA$ and $\cB$ in thermal contact, the total energy $E_{\cA \cB} = E_\cA + E_\cB$ is conserved and the total DOS for given values of $E_\cA$ and $E_\cB$ (assuming weak interaction between the systems) is $\omega_{\cA \cB} = \omega_\cA(E_\cA) \omega_\cB(E_\cB)$. Therefore the maximum entropy $S_{B \cA \cB} = k_B \ln \omega_{\cA \cB} \epsilon$ is obtained by the maximization of $\omega_{\cA \cB}$ with respect to $E_\cA$ and $E_\cB$, under the constrain that $E_{\cA \cB}$ remains constant. This leads to the equilibrium condition for the Boltzmann temperatures, $1/T_{B \cA} \equiv \partial S_{B\cA}/\partial E_\cA = \partial S_{B\cB}/\partial E_\cB \equiv 1/T_{B\cB}$ \cite{AmJPhys.83.163.2015.Frenkel,PhysRevE.92.020103.2015.Swendsen}, so the \textit{postulate 2} is also satisfied. Moreover, once the \textit{postulate 2} is satisfied, under the equilibrium conditions, the \textit{postulate 3} is also satisfied. (We do not discuss the \textit{postulate 4}.)

The definition of $S_B$ is based on the assumption that all the states of the system are equally probable and the system may spend approximately the same amount of time in each of them. In a thermodynamic system the equilibrium state, which corresponds to equal intensive parameters \textit{\'a la} Boltzmann, has an enormous DOS as compared to the non-equilibrium states and for this reason the fluctuations are very small, i.e. the system stays in equilibrium. This assumption is in accordance also with the study of small (mesoscopic) systems where fluctuations are observable (comparable with the averages) and the equilibrium is never achieved in the thermodynamic sense.

The evolution towards equilibrium should be understood in the same sense. If we do not know anything \textit{a priori} about the transition rates between different states, we may assume that they are comparable. Therefore a macroscopic system always evolves towards the parameters regions of higher DOS, i.e. towards equilibrium.

These considerations bring us back to the discussion about the equilibration of the system of spins from Section~\ref{sec_G}. From Eq. (\ref{def_SB}) we obtain $T_B = \omega(E,\bX)/[k_B \nu(E,\bX)]$, where $\nu(E,\bX) \equiv \partial \omega(E,\bX)/\partial E$. If $\nu(E,\bX)<0$, then $T_B(E,\bX)<0$. Therefore for the system of spins described by Eq. (\ref{en_spins}), $T_{Bs}>0$ for $E\in [0,B\mu N_0/2)$, $T_{Bs}<0$ for $E\in (B\mu N_0/2,B\mu N_0]$, and $T_{Bs}(B\mu N_0/2) = \pm\infty$ is undefined. On the other hand, the temperature of the ideal gas $T_{B {\rm id}}$ is positive for any $E$ (and is the same as $T_{G {\rm id}}$ in the thermodynamic limit). Therefore the ideal gas and the system of spins cannot be in thermal equilibrium if $T_{B s} < 0$, contrary to the predictions of Gibbs' formula (\ref{def_SG}), since the system evolves towards parameters corresponding to higher number of states (higher probability).

In general systems of negative $T_B$ cannot be in equilibrium with systems of positive $T_B$ due to the evolution towards maximum DOS and this is in accordance with the experimental observations. The supporters of the Gibbs statistics argue that such situations should not be taken into consideration because the states of negative $T_B$ are metastable.
This is argument is false. The states with negative $T_B$ only appear to be metastable in contact with systems with unbounded spectra.
If in a closed region of space exist only systems of bounded energy spectra, then they may equilibrate at either positive or negative $T_B$ and the systems of negative temperatures would be as legitimate as those of positive temperature. In such a ``world'' one may have reservoirs and thermometers of negative temperature.

\section{Conclusions}

Starting from the Tisza-Callen axiomatic formulation of thermodynamics we analyzed the validity of Gibbs and Boltzmann expressions for the entropy, $S_G$ (\ref{def_SG}) and $S_B$ (\ref{def_SB}), respectively. We agree with the authors of Refs. \cite{AmJPhys.83.163.2015.Frenkel,PhysRevE.92.020103.2015.Swendsen} and disagree with the authors of Refs. \cite{Nat.Phys.10.67.2014.Dunkel, PhysRevE.90.062116.2014.Hilbert, PhysRevE.91.052147.2015.Campisi} in considering $S_B$ as the only generally valid expression for the entropy. $S_G$ is correct only when it gives the same results as $S_B$.


We saw that the equilibrium, according to Boltzmann, is the state of maximum probability (maximum DOS) for the extensive variables. If we assume, by \textit{reductio ad absurdum}, that the real equilibrium state is determined by the Gibbs' prescription and is different from Boltzmann's, then the average value of at least one extensive variable is different from the value corresponding to maximum probability. This implies further that the fluctuations are macroscopic and the equilibrium is not achieved.

We also showed that the negative values of the Boltzmann temperature $T_B$ have clear physical meaning in any statistical ensemble (canonical, microcanonical, etc.). The states of negative $T_B$ seem to be unstable only in the presence of systems of unbounded spectra. If there would be only systems of bounded spectra, then the temperature may take any positive and negative value and one can define thermometers and reservoirs as usual.

\begin{acknowledgement}
This work has been financially supported by CNCSIS-UEFISCDI (project IDEI 114/2011) and ANCS (project PN-09370102). Travel support from Romania-JINR Collaboration grants 05-6-1119-2014/2016, 4436-3-2015/2017, 4342-3-2014/2015, and Titeica-Markov program is gratefully acknowledged.
\end{acknowledgement}

\end{document}